\documentclass[aps,pra,twocolumn,showpacs,preprintnumbers,amsmath,amssymb,superscriptaddress,10pt,nofootinbib]{revtex4-1}

\usepackage{amsmath,amssymb}
\usepackage{mathrsfs}
\usepackage{bm}
\usepackage[latin1]{inputenc} 
\usepackage{dcolumn}
\usepackage{graphicx}
\usepackage{SIunits}
\usepackage{multirow}
\usepackage{ae}
\usepackage{subfig}
\usepackage{hyperref}
\usepackage{color} 
\usepackage{mathtools}

\newcommand{\diff}[0]{\text{d}}

\newcommand{\re}{\mathrm{Re}}
\newcommand{\im}{\mathrm{Im}}

\newcommand{\be}{\begin{equation}}

\begin{document}

\title{Superpositions of Lorentzians  as the class of causal functions}

\author{Christopher A. Dirdal}
\affiliation{Department of Electronics and Telecommunications, Norwegian University of Science and Technology, NO-7491 Trondheim, Norway}

\author{Johannes Skaar}
\affiliation{Department of Electronics and Telecommunications, Norwegian University of Science and Technology, NO-7491 Trondheim, Norway}
\email{johannes.skaar@ntnu.no}

\date{\today}

\begin{abstract}
We prove that all functions obeying the Kramers-Kronig relations can be
approximated as superpositions of Lorentzian functions, to any
precision. As a result, the typical text-book analysis of dielectric
dispersion response functions in terms of Lorentzians may be viewed as
encompassing the whole class of causal functions. A further consequence is that Lorentzian resonances may
be viewed as possible building blocks for engineering any desired
metamaterial response, for example by use of split ring resonators of different parameters. Two example functions, far from typical
Lorentzian resonance behavior, are expressed in terms of Lorentzian
superpositions: A steep dispersion medium that achieves large negative
susceptibility with arbitrarily low loss/gain, and an optimal
realization of a perfect lens over a bandwidth. Error bounds are derived
for the approximation.
\end{abstract}

\maketitle

\section{Introduction}
When considering the dispersion of dielectric media, text book
analysis typically concerns itself with sums of Lorentzian functions \cite{saleh,
jackson, dressel}. While it can be argued that sums of Lorentzians are
physically reasonable response functions for a number of systems,
in particular those described by the Lorentz-model \cite{saleh, jackson,
dressel, caloz11}, it is well known that the only restrictions imposed
by causality are those implied by the Kramers-Kronig relations
\cite{saleh, jackson, dressel,landau_lifshitz_edcm}. In light of the
variety in the electromagnetic responses offered by natural media and
metamaterials, it is of interest to consider the possible gap between
the function space consisting of sums of Lorentzians, and the space
consisting of functions satisfying the Kramers-Kronig relations. To
this end, it is here demonstrated that any complex-valued function
satisfying the Kramers-Kronig relations can be approximated as a
superposition of Lorentzian functions, to any desired accuracy. It therefore follows that the typical analysis of causal behavior in
terms of Lorentzian functions for dielectric or magnetic media encompasses the whole space of functions obeying the
Kramers-Kronig relations. These results therefore serve to
strengthen the generality of the typical analysis of causality.

Two examples where the response functions do not resemble typical
Lorentzian resonance behavior shall here be expressed as Lorentzian
superpositions in order to demonstrate the above findings. Section
\ref{sec:NIES} considers a steep response function which results in a
susceptibility $\re \chi(\omega) \leq -2$ for arbitrarily low loss or
gain \cite{nistad08, Dirdal:13}, and Sec. \ref{sec:perfectlense}
considers an optimal perfect lens response over a bandwidth
\cite{johansen}.  The precisions in both superpositions are shown to
become arbitrarily accurate as the parameters are chosen appropriately.
A natural consequence of such superpositioning is that Lorentzian
functions can be viewed as general building blocks for engineering
causal susceptibilities in metamaterials. Considering that systems such
as the pioneering split-ring resonator implementation \cite{pendry99},
and others \cite{schuller07, cheng2003, caloz11, Shafiei13}, have demonstrated several
 ways of realizing and tailoring Lorentzian responses, this
may prove to be a promising approach.

While literature has so far tended to focus on specific metamaterial
designs, a number of desired responses have emerged for which few
physically viable systems are known \cite{nistad08, johansen, caloz11}.
One such set of response functions are those with desired dispersion
properties \cite{caloz11} which are relevant for applications such as
dispersion compensation \cite{cheng2003, engheta2005}, couplers
\cite{caloz2004}, antenna design \cite{rotman1962, lier2011}, filters
\cite{gil2008}, broadband absorption \cite{feng2012} and broadband
ultra-low refractive index media \cite{Schwartz03}. This leads to the following
question: Starting with a target response, how can one realize an
approximation of it? Towards this end, it has been proposed to use
layered metamaterials \cite{goncharenko07}. Our article considers more generally the
possibility of engineering artificial response functions through the
realization of a finite number of Lorentzians; a method which may be
applicable to a variety of metamaterials. On this note Sec.
\ref{sec:TailorSplitRing} addresses how Lorentzian superposition responses 
can be realized through the arrangement of split ring cylinders of different radii and material parameters. Finally Sec.
\ref{sec:Errorestimate} derives an estimate of the error that arises when
a target response is approximated by a finite sum of Lorentzian functions.

The following section will set out the main results of this article
while leaving detailed calculations to later sections and appendices.

\section{Superposition of Lorentzians}
A Lorentzian function can be written in the form
\begin{equation}
L(\omega)=\frac{2}{\pi}\frac{\omega_0}{\omega_0^2-\omega^2-i\omega\Gamma}, 
\label{eq:Lorentz}
\end{equation} where $\omega_0\geq 0$ is the resonance frequency, $\omega$ is the frequency, and $\Gamma \geq 0$ is the bandwidth. It may be demonstrated that the imaginary part of $L(\omega)$ approaches a sum of two Dirac delta-functions with odd symmetry as
\begin{equation}
\im L(\omega) \overset{}{\longrightarrow} \delta(\omega - \omega_0) - \delta(\omega + \omega_0), \label{eq:LorentzToTwoDeltas}
\end{equation} when $\Gamma\to 0$. This is exemplified in Fig. \ref{fig:LorConvergence} and proven in Appendix \ref{sec:DeltaConv}. A goal function, such as the imaginary part of a susceptibility $\im \chi(\omega)$, may therefore be expressed
\begin{equation} 
\im\chi(\omega)=\frac{2}{\pi}\lim_{\Gamma\to 0}\int_0^\infty\im\chi(\omega_0) \im \bigg \{ \frac{\omega_0}{\omega_0^2-\omega^2-i\omega\Gamma} \bigg \}\diff\omega_0 .
\label{eq:ImChi}
\end{equation} This limit integral expression may then be approximated by the sum
\begin{figure}[hb]
  \centering
  \includegraphics[width=0.44\textwidth]{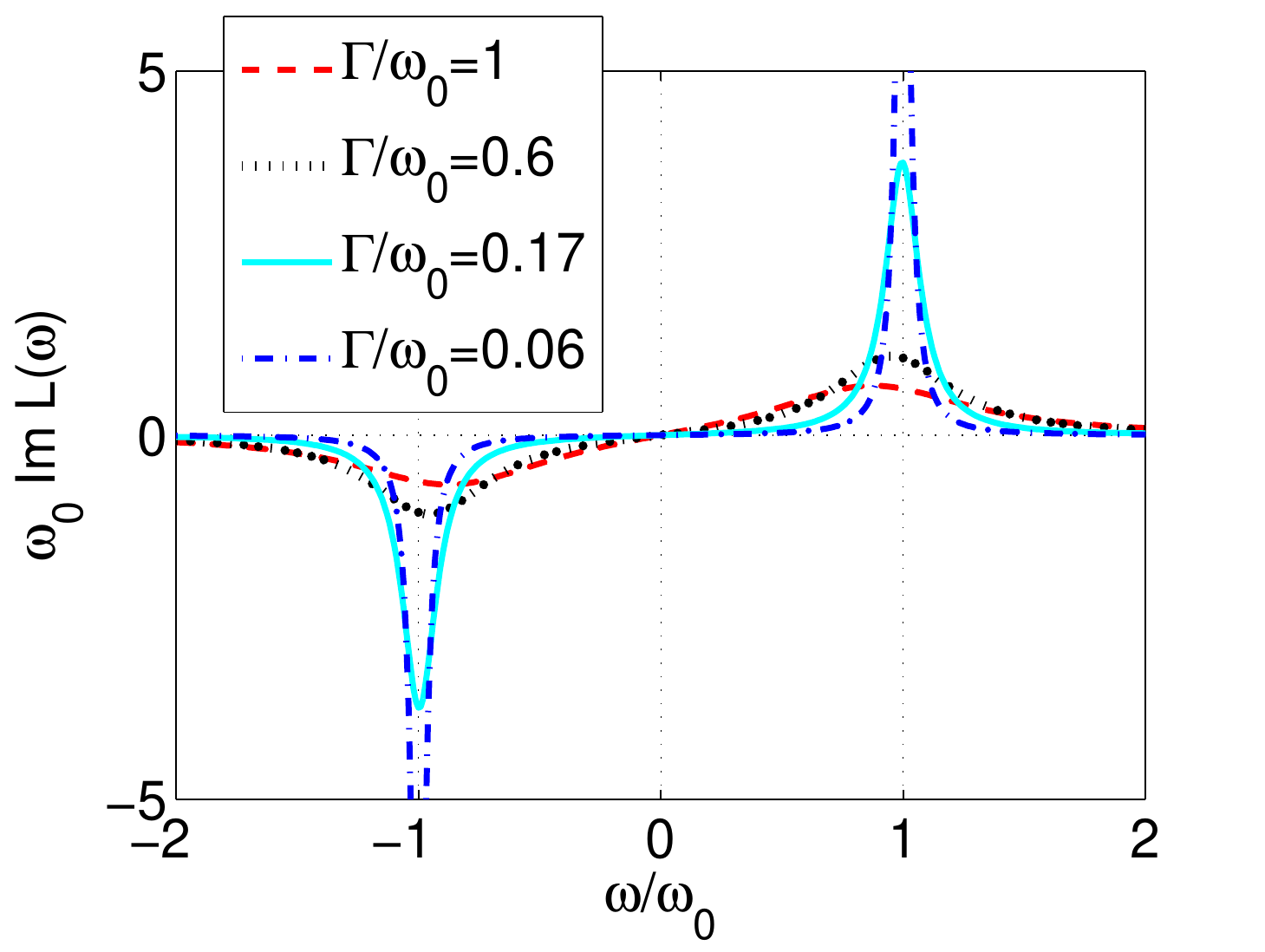}
  \caption{Imaginary part of the Lorentzian function $L(\omega)$ \eqref{eq:Lorentz} approaching a sum of two delta-functions for decreasing values of $\Gamma$.}
  \label{fig:LorConvergence}
\end{figure}
\begin{equation}\label{representionsum}
\im \chi(\omega) \approx\frac{2}{\pi}\sum_{m=0}^{M-1}\im\chi(\omega_m) \im \bigg \{ \frac{\omega_m\Delta}{\omega_m^2-\omega^2-i\omega\Gamma} \bigg \} ,
\end{equation} where $\omega_m=\Delta/2+m\Delta$, $\Delta$ is the resolution along the integration variable $\omega_0$, and $M$ is a large integer. This sum, which shall be designated $\im \chi_{\Gamma, \Delta}(\omega)$, can be made to approximate $\im \chi(\omega)$ to any desired degree of accuracy. More precisely, the involved error is shown to converge to zero in both $L^\infty$ and $L^2$:
\begin{equation}
\big \| \im \chi_{\Gamma, \Delta}(\omega) - \im \chi(\omega) \big \| \to 0 \quad \text{as} \quad \Gamma \to 0, \label{eq:L2ImChiNew}
\end{equation} where $\Delta$ is chosen suitably, e.g. $\Delta = \Gamma / \sqrt{M}$ (Appendix \ref{sec:L2Conv}). Considering \eqref{representionsum}, one may define a function
\begin{equation} \label{representionsum2}
\chi_{\Gamma, \Delta}(\omega) = \frac{2}{\pi}\sum_{m=0}^{M-1}\im\chi(\omega_m) \frac{\omega_m\Delta}{\omega_m^2-\omega^2-i\omega\Gamma},
\end{equation} where now both real and imaginary parts of the Lorentzians are superposed. From the Kramers-Kronig relations one then has
\begin{equation}
-\mathscr{H}\big (\im \chi(\omega) - \im \chi_{\Gamma, \Delta}(\omega) \big ) = \re \chi (\omega) - \re \chi_{\Gamma, \Delta}(\omega)
\end{equation} where $\mathscr{H}$ represents the Hilbert transform and $\re \chi_{\Delta, \Gamma}(\omega)$ the real part of \eqref{representionsum2}. Since the Hilbert transform preserves the energy, or $L^2$-norm, it therefore follows that
\begin{equation}
\big \| \re \chi_{\Gamma, \Delta}(\omega) - \re \chi(\omega) \big \| \to 0  \quad \text{as} \quad \Gamma \to 0, \label{eq:L2ReChiNew}
\end{equation} when given \eqref{eq:L2ImChiNew}. On the basis of this, it follows that
\begin{equation}
\chi(\omega) \approx \chi_{\Gamma, \Delta}(\omega), \label{eq:centralResult1}
\end{equation} meaning that both real and imaginary parts of $\chi(\omega)$ are approximated by the summation of Lorentzians. The terms are weighted by $\im \chi (\omega)$ at each resonance frequency according to \eqref{representionsum2}. Eq. \eqref{eq:centralResult1} combined with \eqref{representionsum2} therefore becomes the central result of this article. Noting that \eqref{representionsum2} is itself a Riemann sum, it follows that the limit integral expression
\begin{equation}\label{represention}
\chi(\omega)=\frac{2}{\pi}\lim_{\Gamma\to 0}\int_0^\infty\im\chi(\omega_0) \frac{\omega_0}{\omega_0^2-\omega^2-i\omega\Gamma}\diff\omega_0,
\end{equation} may be written on the basis of the preceding arguments.  Eq. \eqref{representionsum2} and \eqref{represention} therefore approximate the space of functions satisfying the Kramers-Kronig relations as superpositions of Lorentzians to any degree of precision. In the following section this result shall be demonstrated with two examples.



The validity of \eqref{eq:L2ImChiNew} requires that $\chi(\omega)$ is analytic on the real axis (not only in the upper half-plane). In the event of non-analytic susceptibilities on the real axis, however, all problems are bypassed by instead evaluating $\im \chi(\omega)$ along the line $\omega+i\delta$ before approximating by Lorentzians. Here $\delta>0$ is an arbitrarily small parameter. Since $\chi(\omega)$ is analytic there, \eqref{eq:L2ImChiNew} is valid. Furthermore, since $\im \chi(\omega+i\delta)\to \im\chi(\omega)$ almost everywhere as $\delta \to 0$, the representation can be made arbitrarily accurate, meaning that any $\im \chi(\omega)$ can be approximated to any precision. In fact, this also includes media that do not strictly obey the Kramers-Kronig relationship due to singularities on the real axis, such as the ideal plasma.
 


\section{Examples}
\label{sec:Examples}

\subsection{Susceptibility where $\re \chi \leq -2$ through a steep response}
\label{sec:NIES}
It is possible to achieve $\re \chi \leq -2$ with arbitrarily low loss or gain \cite{nistad08, Dirdal:13}. Consider a susceptibility with
\begin{equation}
\im \chi (\omega) = \begin{cases} \omega/\omega_c &\mbox{if } |\omega| < \omega_c \\
0 & \mbox{elsewhere }. \end{cases} \label{eq:NIES}
\end{equation} As a result of the infinite steepness at $\omega=\omega_c$, the Hilbert transform gives $\re \chi (\omega_c) = - \infty$. It follows that it is possible to scale \eqref{eq:NIES} to make the magnitude $|\im \chi(\omega)|$ arbitrarily small for all frequencies while maintaining $\re \chi(\omega_c) \leq -2$. 

\begin{figure}[htb]
\centering
\subfloat[Real and imaginary parts of the integral and sum approximations by \eqref{represention} and \eqref{representionsum2}, respectively. The drops above the frequency axis correspond to the imaginary parts, whereas the dips below the frequency axis correspond to the real parts.]{\label{fig:NIES}\includegraphics[width=0.44\textwidth]{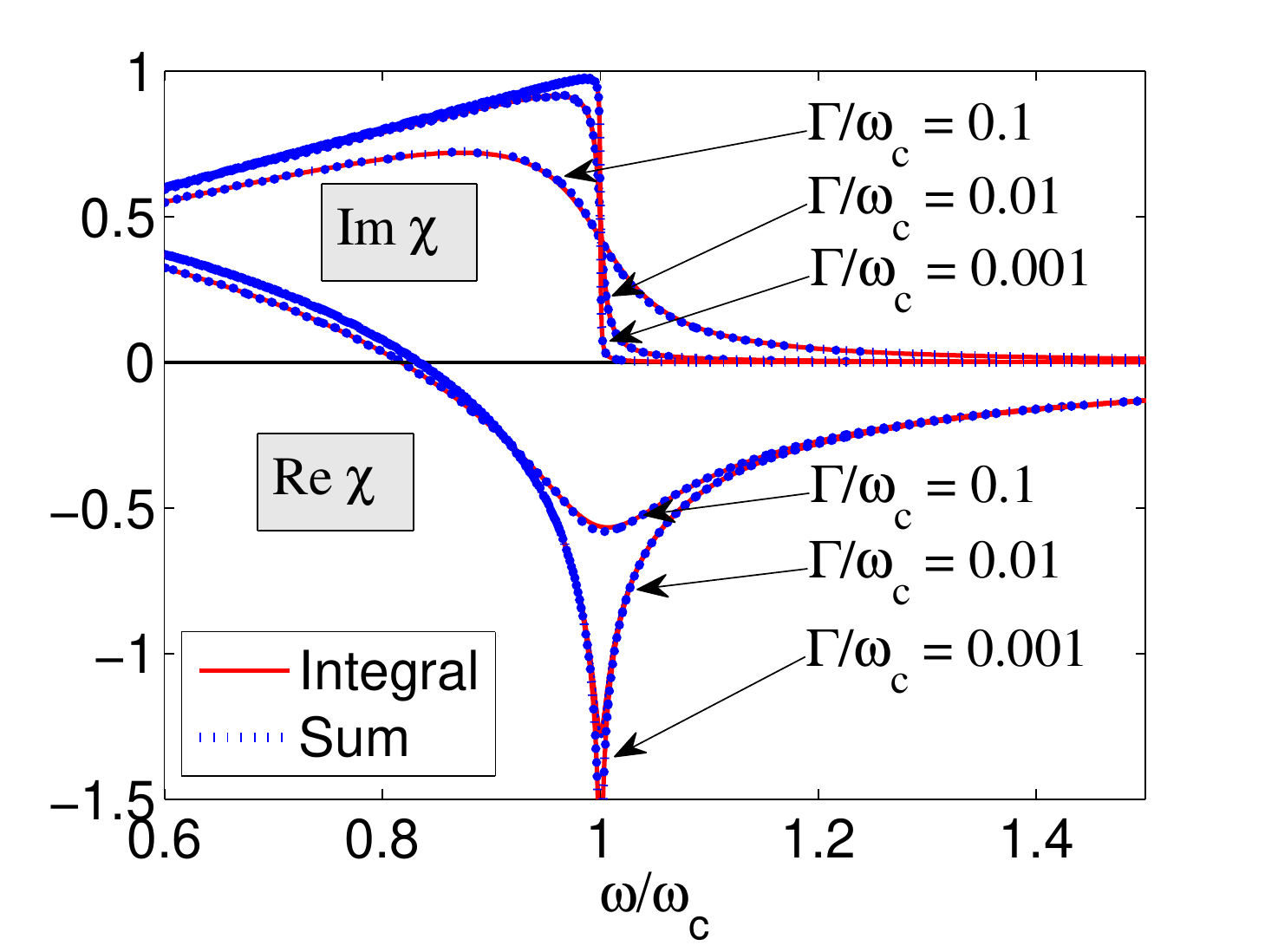}} \\
\subfloat[(Color online) Displaying the imaginary parts for one out of every 5 Lorentzians in the sum \eqref{representionsum2}, where $\Gamma/\omega_c=0.01$ and $\Delta/\omega_c=0.005$.]{\label{fig:SummedLor}\includegraphics[width=0.44\textwidth]{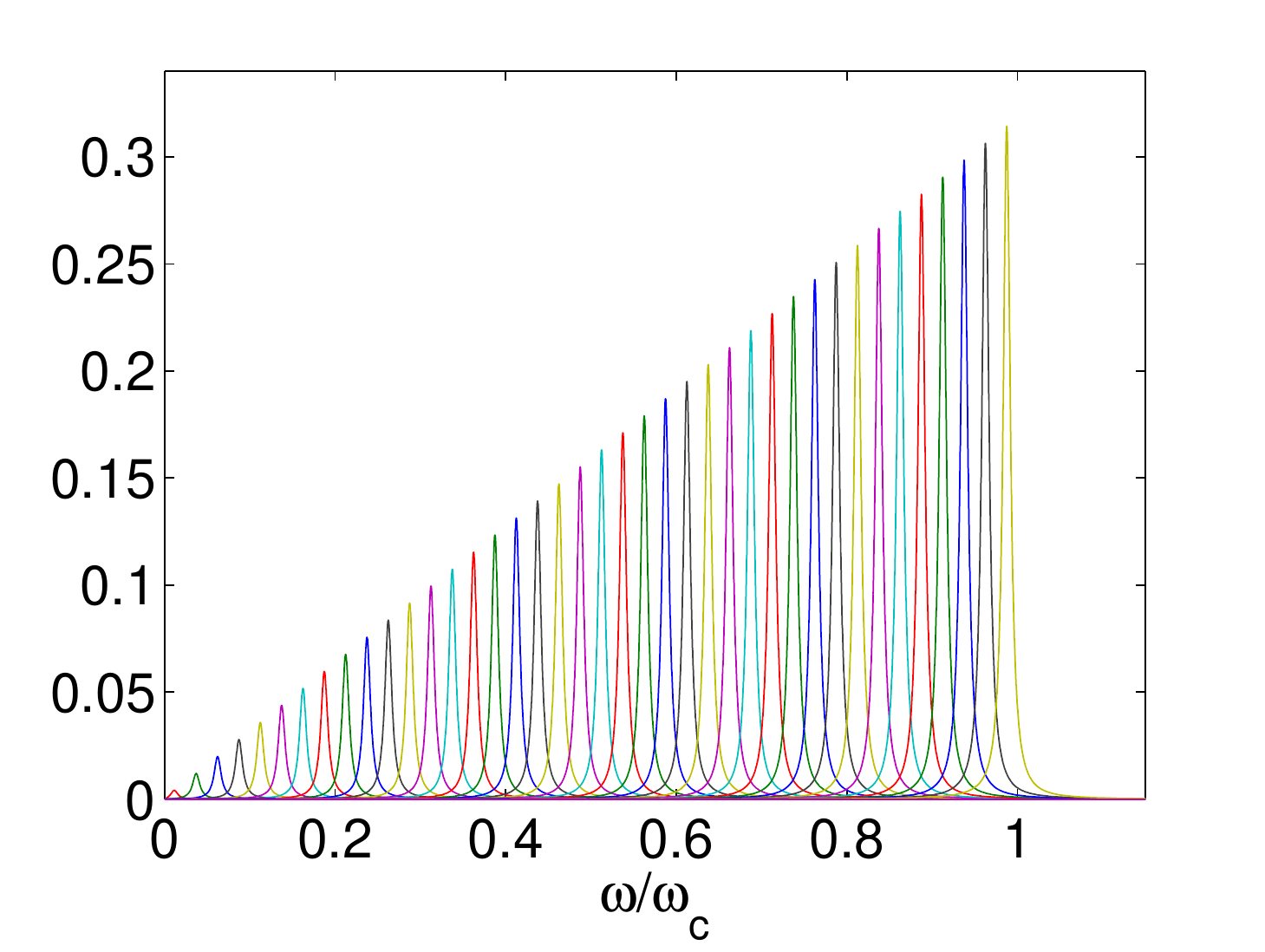}} 
\caption{}
\label{fig:Example1}
\end{figure}

Inserting \eqref{eq:NIES} as $\im \chi(\omega_0)$ in \eqref{represention} gives
\begin{eqnarray}
\chi ( \omega) =\frac{2}{\pi} \lim_{\Gamma \to 0} \bigg [ 1 - &i& \sqrt{\bigg (\frac{\omega}{\omega_c}\bigg )^2 + i \frac{\omega \Gamma}{\omega_c^2}} \nonumber \\
 &&\arctan \bigg ( \frac{1}{i \sqrt{ (\omega/\omega_c)^2 + i (\omega \Gamma/\omega_c^2)}} \bigg ) \bigg ]. \nonumber \\ \label{eq:NIESInt}
\end{eqnarray} One can show that as $\Gamma\to 0$, the imaginary part of \eqref{eq:NIESInt} yields \eqref{eq:NIES}. Using \eqref{representionsum2} one may likewise approximate the response \eqref{eq:NIES} as a finite sum of Lorentzians: Fig. \ref{fig:NIES} plots the real and imaginary parts of $\chi(\omega)$ as approximated by both sum and integral expressions, \eqref{representionsum2} and \eqref{represention} respectively, where $\Gamma/\omega_c$ is chosen equal to $0.1$, $0.01$, and $0.001$, and where $\Delta= \Gamma/ 2$. One observes that the sum \eqref{representionsum2} falls in line with the integral result \eqref{eq:NIESInt}. For $\Gamma/\omega_c=0.001$ one has $\re \chi(\omega_c)= -2$. As $\Gamma \to 0$, meaning that the drop in the approximated curves of $\im \chi(\omega)$ at $\omega/\omega_c=1$ become infinitely steep, one has that $\re \chi(\omega_c) \to -\infty$ in both cases.

The imaginary parts for one out of every 5 Lorentzians in the sum \eqref{representionsum2} are displayed in Fig. \ref{fig:SummedLor} for $\Gamma/\omega_c = 0.01$. 

\subsection{Perfect lens}
\label{sec:perfectlense}
\begin{figure}[htb]
\centering
\subfloat[System designed to minimize $|\chi + 2|$ over a bandwidth. The red solid curve represents $\im \chi$ whereas the blue dashed curve represents $\re \chi$.]{\label{fig:PerfLens}\includegraphics[width=0.44\textwidth]{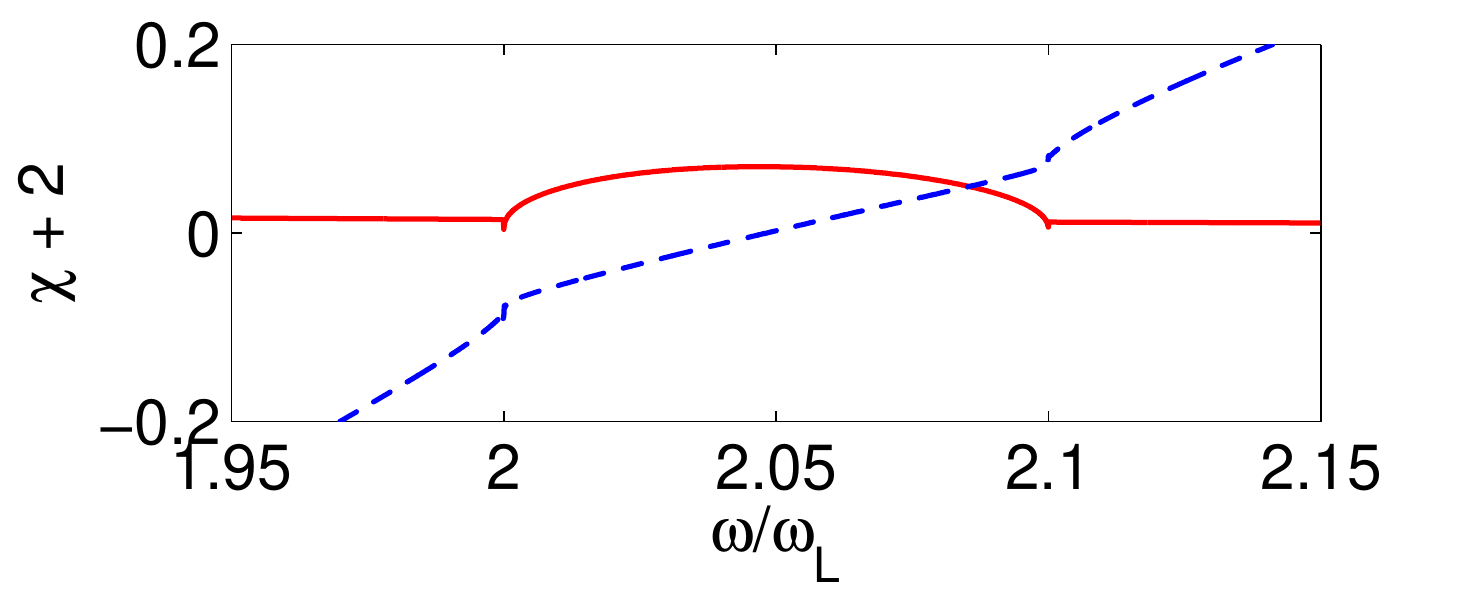}} \\
\subfloat[Sum \eqref{representionsum2} of Lorentzians over the interval $\omega/ \omega_L \in (1.95, 2.15)$, where $\Gamma / \omega_L=0.001$ and $\Delta/\omega_L=5 \cdot 10^{-4}$.]{\label{fig:PerflensSum}\includegraphics[width=0.44\textwidth]{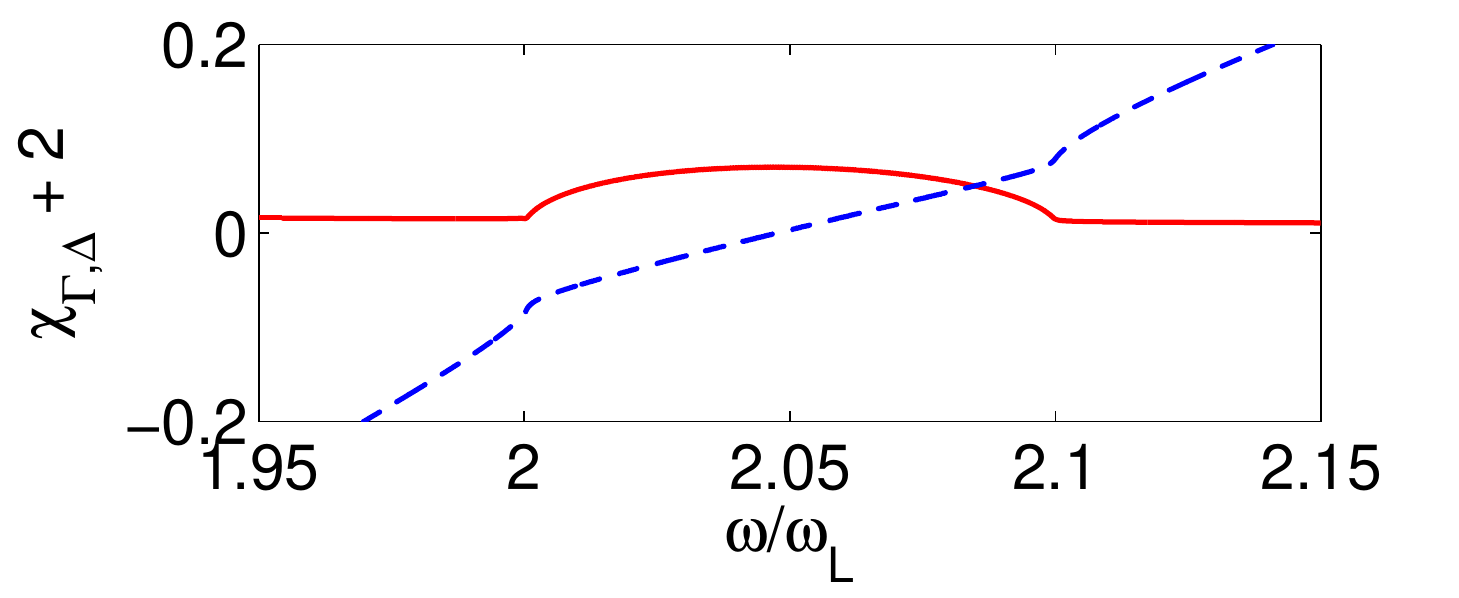}} \\
\subfloat[Plot of $|\chi + 2|$ for both the system in (a) (dashed black curve) as well as two sum approximations. The solid curve corresponds with the absolute value of (b).]{\label{fig:AbsPerflensSum}\includegraphics[width=0.44\textwidth]{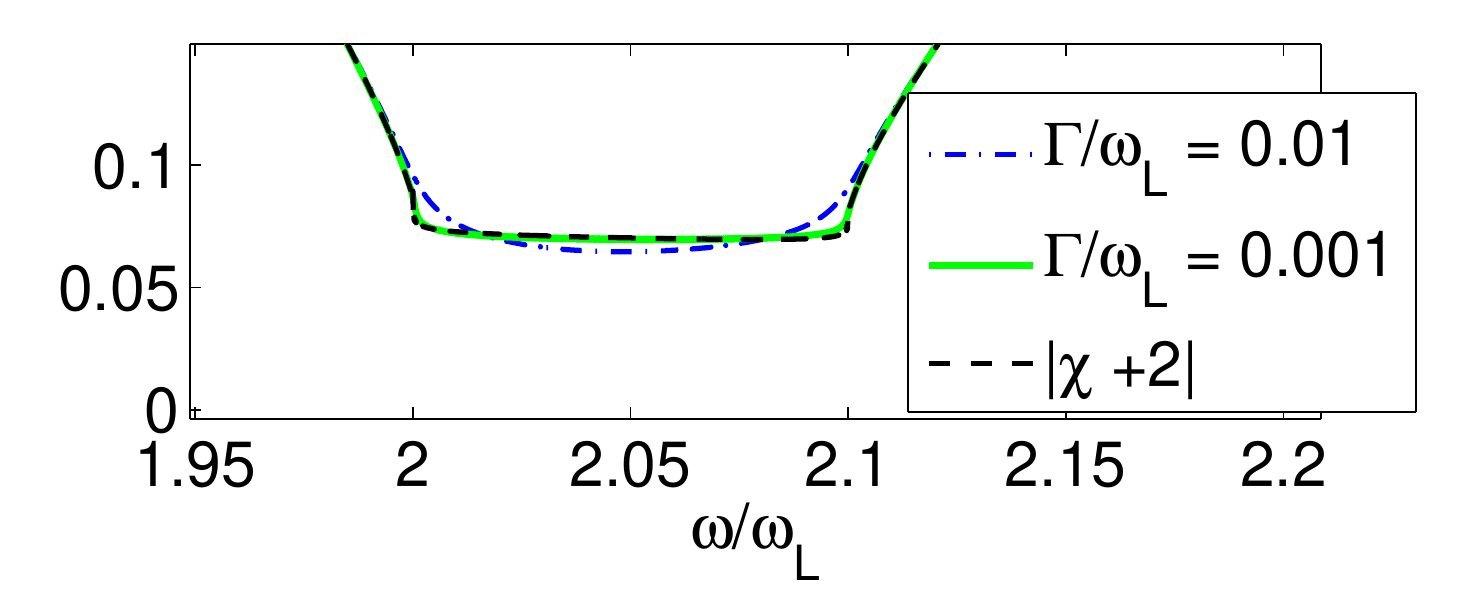}} \\
\subfloat[(Color online) The imaginary parts for one out of every five Lorentzians in the sum displayed in (b).]{\label{fig:PerflensSumComp}\includegraphics[width=0.44\textwidth]{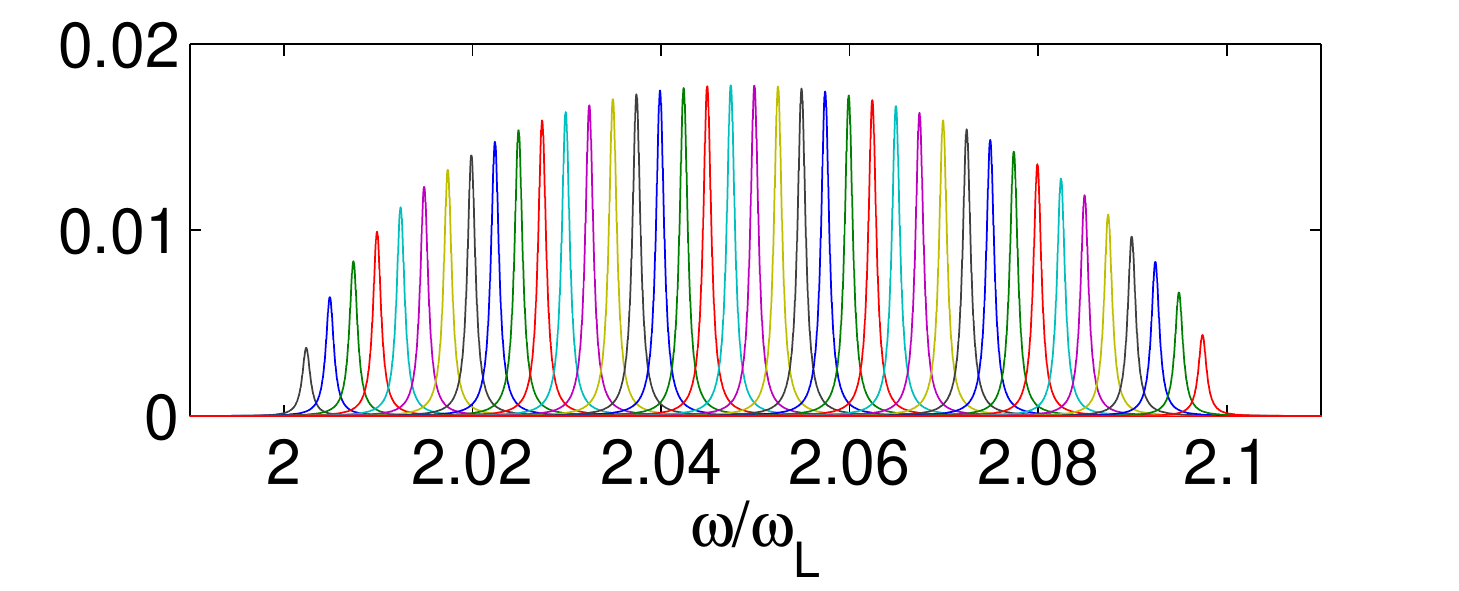}} 
\caption{Perfect lens}
\label{fig:Example2}
\end{figure}

It has been shown in \cite{ramakrishna2002,johansen} that the resolution, $r$, for a metamaterial lens of thickness $d$ surrounded by vacuum, is given by
\begin{eqnarray}
r = \frac{-2 \pi d}{\ln \big (|\chi(\omega) +2|/2\big )}, \label{eq:Resolution}
\end{eqnarray} for a one dimensional image object. Here, $\chi(\omega)$ is either the electric or magnetic susceptibility depending on the polarization of the incident field. It follows that any perfect lens should approximate $|\chi + 2|=0$ over a bandwidth. A system approaching such an optimum is displayed in Fig. \ref{fig:PerfLens} \cite{johansen}. There has been placed a strong Lorentzian resonance at $\omega=\omega_L$ (out of view) and a weak, slowly varying function around $\omega/\omega_L \sim 2.05 $. Taking the absolute value gives the black dashed curve in Fig. \ref{fig:AbsPerflensSum}, which reveals that $|\chi + 2|$ remains small and constant over the bandwidth.

Fig. \ref{fig:PerflensSum} represents a sum \eqref{representionsum2} of Lorentzians over the interval $\omega/ \omega_L \in (1.95, 2.15)$ where $\Gamma / \omega_L=0.001$ and $\Delta/\omega_L =5 \cdot 10^{-4}$. The goal function for \eqref{representionsum2} has been found by subtracting the strong resonance situated at $\omega=\omega_L$ from $\im \chi(\omega)$ in Fig. \ref{fig:PerfLens}. The strong resonance has then later been added to the sum of Lorentzians \eqref{representionsum2}. The absolute value of the real and imaginary parts in Fig. \ref{fig:PerflensSum} correspond with the green solid curve in Fig. \ref{fig:AbsPerflensSum}, which neatly follows the dashed curve of $|\chi + 2|$. Also observed in Fig. \ref{fig:AbsPerflensSum} is the corresponding result of another sum approximation where the Lorentzians are wider. Hence one observes the trend that as $\Gamma$ decreases the Lorentzian sum \eqref{representionsum2} approximates $|\chi + 2|$ increasingly well. 

Fig. \ref{fig:PerflensSumComp} displays the imaginary parts for one out of every five Lorentzians in the sum corresponding with Fig. \ref{fig:PerflensSum}.

\section{Lorentzian functions as building blocks for tailoring causal susceptibilities}
\label{sec:BuildingBlocks}

\subsection{Tailoring responses in split ring resonators}
\label{sec:TailorSplitRing}
Sections \ref{sec:NIES} and \ref{sec:perfectlense} have demonstrated that useful responses which do not arise in any natural systems can nevertheless be approximated as a superposition of ordinary Lorentzian resonances to any desired precision. Considering that Lorentzian resonances can be realized and tailored for both permittivities $\epsilon$ and permeabilities $\mu$ in numerous metamaterial realizations \cite{schuller07, pendry99, caloz11, Shafiei13}, this is of particular interest for the prospect of engineering desired metamaterial responses. For instance, in the case of an array of split ring cylinders one may find for the effective permeability \cite{pendry99}: 
\begin{eqnarray}
\mu_\text{eff} -1 = \frac{\omega^2 F}{\frac{2}{\pi^2 \mu_0 C r^3} -\omega^2 - i \omega \frac{2 \rho }{r \mu_0}}, \label{eq:PendryMu}
\end{eqnarray} where $C$ is the capacitance per $\text{m}^2$ in the split ring cylinder, $r$ is its radius, $\rho$ is the resistance per circumference-length ratio and $F$ is the fractional volume of the unit cell occupied by the interior of the cylinder. For small bandwidths the presence of $\omega^2$ in the numerator is not significant and \eqref{eq:PendryMu} approximates a Lorentzian response. By comparing it with \eqref{eq:Lorentz} one may determine the Lorentzian parameters as
\begin{eqnarray}
\omega_0^2 &=& \frac{2}{\pi^2 \mu_0 C r^3} \\
\Gamma &=& \frac{2 \rho }{r \mu_0}
\end{eqnarray} The resonator strength $F$ is expressed
\begin{eqnarray}
F &=& \frac{\pi r^2}{a^2},
\end{eqnarray} where $a$ is the dimension of the unit cell. Hence the resonance frequencies, widths, and strengths can be tailored by varying $r$, $\rho$, $C$. It may be noted that split ring cylinders display large resistivity in the optical regime, making it difficult to achieve narrow Lorentzian responses there. For this reason other systems, such as metamolecules of nanoparticles, have been proposed for optical purposes \cite{Shafiei13}.

\begin{figure}[]
\centering
\def\svgwidth{0.5\columnwidth}
\begingroup%
  \makeatletter%
  \providecommand\color[2][]{%
    \errmessage{(Inkscape) Color is used for the text in Inkscape, but the package 'color.sty' is not loaded}%
    \renewcommand\color[2][]{}%
  }%
  \providecommand\transparent[1]{%
    \errmessage{(Inkscape) Transparency is used (non-zero) for the text in Inkscape, but the package 'transparent.sty' is not loaded}%
    \renewcommand\transparent[1]{}%
  }%
  \providecommand\rotatebox[2]{#2}%
  \ifx\svgwidth\undefined%
    \setlength{\unitlength}{148.625bp}%
    \ifx\svgscale\undefined%
      \relax%
    \else%
      \setlength{\unitlength}{\unitlength * \real{\svgscale}}%
    \fi%
  \else%
    \setlength{\unitlength}{\svgwidth}%
  \fi%
  \global\let\svgwidth\undefined%
  \global\let\svgscale\undefined%
  \makeatother%
  \begin{picture}(1,1.11714446)%
    \put(0,0){\includegraphics[width=\unitlength]{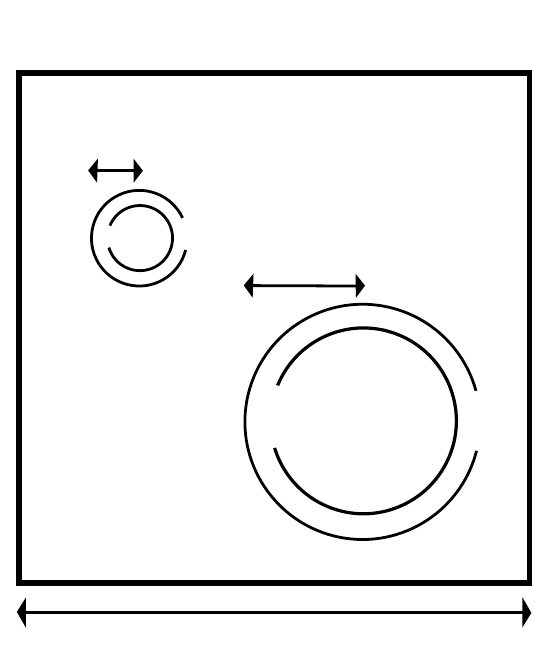}}%
    \put(0.18851057,0.95124317){\color[rgb]{0,0,0}\makebox(0,0)[lb]{\smash{$r_1$}}}%
    \put(0.53739076,0.72547227){\color[rgb]{0,0,0}\makebox(0,0)[lb]{\smash{$r_2$}}}%
    \put(0.48,0.00665475){\color[rgb]{0,0,0}\makebox(0,0)[lb]{\smash{$a$}}}%
  \end{picture}%
\endgroup%
\caption{Base of a unit cell with height $l$ containing two different cylinders.}
\label{fig:unitCell}
\end{figure}
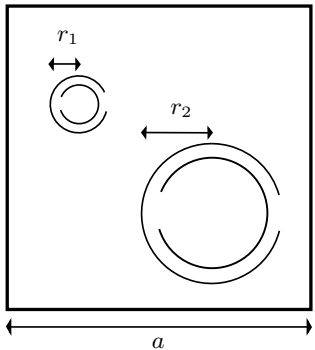

In order to realize a sum of Lorentz resonances such as those leading to Fig. \ref{fig:SummedLor} and Fig. \ref{fig:PerflensSumComp} by means of split ring cylinders, one could propose to place different cylinders in each unit cell. The idea would be to realize and superpose a number of different Lorentzian resonances of the form \eqref{eq:PendryMu} corresponding with different values of $r$, $\rho$, $C$. The density of each type of split ring cylinder can then be thought to give the appropriate resonance strength. However, as it is known that two dipoles in close vicinity influence each others dipole moment, it is not intuitively clear that the total response will be as simple as a superposition of the individual responses. The remainder of this section will therefore be used to demonstrate this. Towards this end, the procedure outlined in \cite{pendry99} will be modified to calculate the effective permeability $\mu_\text{eff}$ of a split ring cylinder metamaterial when every unit cell contains two split ring cylinders of different radii (Fig. \ref{fig:unitCell}).

The effective permittivity $\mu_\text{eff}$ is expressed by finding the effective macroscopic fields $B_\text{ave}$ and $H_\text{ave}$ over the array from the corresponding actual fields $B$ and $H$ in each unit cell:
\begin{eqnarray}
\mu_\text{eff} &=& B_\text{ave} / \mu_0 H_\text{ave}. \label{eq:muEff}
\end{eqnarray} One may show that 
\begin{eqnarray}
B_\text{ave}&=& \mu_0 H_0 \\
H_\text{ave}&=& H_0 - \frac{1}{a^2} \big ( \pi r_1^2 j_1 + \pi r_2^2 j_2\big ) , \label{eq:H_ave}
\end{eqnarray} where $j_{1,2}$ is the surface current per unit length along the circumference of the cylinders with radii $r_1$ and $r_2$ respectively. When finding an expression for the field inside each cylinder $H_1$ and $H_2$, \eqref{eq:H_ave} is used to express
\begin{eqnarray}
H_{1,2}= H_\text{ave} + j_{1,2}. \label{eq:fieldInEachCylinder}
\end{eqnarray} Since both split ring cylinders observe the same interaction field contained in $H_\text{ave}$, application of Faraday's law to each cylinder leads to two equations that may be solved individually for $j_1$ and $j_2$:
\begin{eqnarray}
j_{1,2}= \frac{-H_\text{ave}}{1 + \frac{i 2 \rho_{1,2}}{\omega r_{1,2} \mu_0} - \frac{2}{\mu_0 \omega^2 \pi^2 r_{1,2}^3 C_{1,2} } }. \label{eq:SurfaceCurrents}
\end{eqnarray} In order to evaluate \eqref{eq:muEff}, one may now rearrange \eqref{eq:H_ave} to find an expression for $H_0$ in terms of $j_1$ and $j_2$, for which one in turn can use \eqref{eq:SurfaceCurrents} to find
\begin{eqnarray}
\mu_\text{eff} -1= \sum_k \frac{ \omega^2 F_k}{ \frac{2}{\pi^2 \mu_0  C_{k} r_{k}^3 } -\omega^2  -i \omega \frac{ 2 \rho_{k}}{r_{k} \mu_0}}. \label{eq:effectiveMusrc}
\end{eqnarray} Here $k=\{1,2 \}$ and $F_k= \pi r_k^2 /a^2$ is the volume fraction occupied by each split ring cylinder. By comparison with \eqref{eq:PendryMu} one observes that the resulting $\mu_\text{eff} - 1$ here is simply the superposition of the individual split ring cylinder responses as found in \cite{pendry99}. This comes as a consequence of the interaction field sensed by both split ring cylinders being uniform. This description of the interaction is valid as long as the cylinders are long, for which the returning field lines at the end of each cylinder are spread uniformly over the unit cell. 

Finally, the analysis presented here is easily generalized for unit cells with many different split ring cylinders, for which \eqref{eq:effectiveMusrc} remains valid and $k$ labels the different split ring cylinders in the unit cell.

\subsection{Error estimate}
\label{sec:Errorestimate}

Towards the goal of engineering a desired response $\chi(\omega)$ by use of a finite number of Lorentzians with non-zero widths ($\Gamma > 0$), it is of interest to quantify the precision of such an approximation. An error estimate can be found by considering the combined error introduced in both the delta-function approximation \eqref{eq:ImChi} of $\im \chi(\omega)$ for finite $\Gamma$, and the Riemann sum approximation \eqref{representionsum} of the integral \eqref{eq:ImChi}. The total error may therefore be bounded according to
\begin{eqnarray}
|\im \chi(\omega) - \im \chi_{\Gamma, \Delta}(\omega)| &\leq &|\im \chi_\Gamma(\omega) - \im \chi (\omega)| \nonumber \\
&+& |\im \chi_{\Gamma, \Delta}(\omega) - \im \chi_\Gamma(\omega)| , \label{eq:TotalError}
\end{eqnarray} where $\im \chi_\Gamma (\omega)$ represents \eqref{eq:ImChi} without the limit and $\im \chi_{\Gamma, \Delta}(\omega)$ represents the sum \eqref{representionsum}. In what follows, bounds for the Riemann sum approximation error $ |\im \chi_{\Gamma, \Delta}(\omega) - \im \chi_\Gamma(\omega)|$ and delta-convergence error $|\im \chi_{\Gamma}(\omega) - \im \chi_\Gamma(\omega)|$ shall be derived in turn.

\subsubsection{Riemann sum approximation error}
Defining $\omega_M=M\Delta$, and naming  the integrand in \eqref{eq:ImChi} $f_\text{int}$, one may find an upper bound on the Riemann approximation error to be
\begin{align}\label{errestimation}
&|\im \chi_\Gamma(\omega)- \im \chi_{\Gamma,\Delta}(\omega)| \leq \max{|f_\text{int}''|}\frac{\omega_M^3}{24M^2}+|\chi_{\text{err}}(\omega)| \nonumber\\
&< |\im \chi(\omega)|  \frac{\omega_M\Delta^2}{3\Gamma^3} \left(1 + O\left(\frac{\Gamma^2}{\omega^2}\right) \right)+ |\chi_{\text{err}}(\omega)|,
\end{align}
where
\begin{equation}\label{errint}
\chi_{\text{err}}(\omega)=\frac{2}{\pi}\int_{\omega_M}^\infty\im\chi(\omega_0) \im \bigg \{ \frac{\omega_0}{\omega_0^2-\omega^2-i\omega\Gamma} \bigg \} \diff\omega_0 .
\end{equation} In obtaining the last inequality in \eqref{errestimation}, it has been used that $f_\text{int}''$ is dominated by the curvature of the Lorentzian for sufficiently small $\Gamma$. After some algebra one finds $|f_\text{int}''|\leq |\im \chi(\omega)| 8/\Gamma^3( 1 + O(\Gamma^2/\omega^2))$ when assuming that $\Gamma / \omega \ll 1$. Note that \eqref{errint} can be rewritten
\begin{eqnarray}\label{errint2}
\chi_\text{err}(\omega)=\frac{2}{\pi}\int_{0}^\infty  &\bigg [& 1 - \text{rect} \bigg (\frac{\omega_0}{\omega_M} \bigg) \bigg ] \im\chi (\omega_0) \nonumber \\
&& \im \bigg \{ \frac{\omega_0}{\omega_0^2-\omega^2-i\omega\Gamma} \bigg \} \diff\omega_0.
\end{eqnarray} By taking the limit $\Gamma \to 0$ this expression becomes of the form \eqref{eq:ImChi}, and hence the definition of delta convergence can be used to find the limit of $\chi_\text{err}(\omega)$:
\begin{eqnarray} \label{eq:deltConvResult}
\lim_{\Gamma \to 0} \chi_\text{err}(\omega) =  \bigg [  1 - \text{rect} \bigg (\frac{\omega}{\omega_M} \bigg)  \bigg ] \im \chi(\omega).
\end{eqnarray} It is observed that $|\chi_\text{err}(\omega)|$ vanishes as $\omega_M \to \infty$. For $\omega_M / \Gamma \gg 1$ the error involved in replacing $\chi_\text{err}(\omega)$ by its limiting value \eqref{eq:deltConvResult} becomes small. Therefore, for most goal functions $\im \chi(\omega)$, where the sum \eqref{representionsum2} covers the frequency interval of interest, the error described by $|\chi_\text{err}(\omega)|$ is roughly as small as $|\im \chi(\omega)|$ outside of this interval.

\subsubsection{Delta-convergence approximation error}
\label{sec:DeltaConvApprozError}
The delta-convergence approximation error 
\begin{equation}
\psi_\text{err}(\omega) = |\im \chi_\Gamma(\omega) - \im \chi (\omega)|, \label{eq:psiErrDef}
\end{equation} which arises when using Lorentzians of finite width ($\Gamma>0$) rather than delta-functions in \eqref{eq:ImChi}, shall now be evaluated. Consider Fig. \ref{fig:DeltaKonv}: The imaginary part of an arbitrary response $\chi(\omega)$ is to be approximated according to \eqref{eq:ImChi} with finite $\Gamma$, and the positive frequency peak of the imaginary part of a Lorentzian $\im L_{+}(\omega,\omega_0)$ is displayed. The Lorentzian \eqref{eq:Lorentz} can be expanded in terms of partial fractions as:
\begin{eqnarray}
\frac{2}{\pi} \im \bigg \{ \frac{\omega_0}{\omega_0^2 - \omega^2 - i \omega \Gamma} \bigg \} = \im L_{+}(\omega,\omega_0) - \im L_{-}(\omega,\omega_0), \label{eq:ExpressLorthruF2} \nonumber \\ 
\end{eqnarray} where $\im L _{\pm}(\omega,\omega_0)$ corresponds to positive and negative frequency peaks of the Lorentzian respectively, according to:
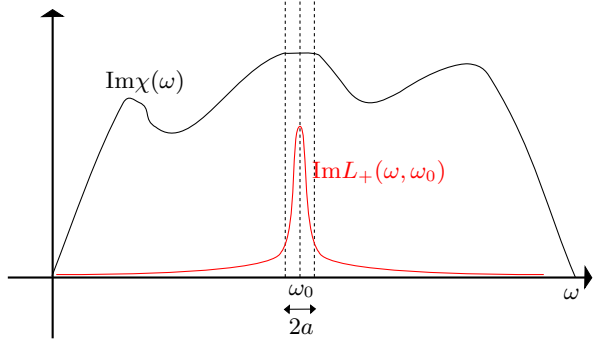
\begin{figure}[]
\centering
\def\svgwidth{0.9\columnwidth}
\begingroup%
  \makeatletter%
  \providecommand\color[2][]{%
    \errmessage{(Inkscape) Color is used for the text in Inkscape, but the package 'color.sty' is not loaded}%
    \renewcommand\color[2][]{}%
  }%
  \providecommand\transparent[1]{%
    \errmessage{(Inkscape) Transparency is used (non-zero) for the text in Inkscape, but the package 'transparent.sty' is not loaded}%
    \renewcommand\transparent[1]{}%
  }%
  \providecommand\rotatebox[2]{#2}%
  \ifx\svgwidth\undefined%
    \setlength{\unitlength}{483.61601563bp}%
    \ifx\svgscale\undefined%
      \relax%
    \else%
      \setlength{\unitlength}{\unitlength * \real{\svgscale}}%
    \fi%
  \else%
    \setlength{\unitlength}{\svgwidth}%
  \fi%
  \global\let\svgwidth\undefined%
  \global\let\svgscale\undefined%
  \makeatother%
  \begin{picture}(1,0.52763968)%
    \put(0,0){\includegraphics[width=\unitlength]{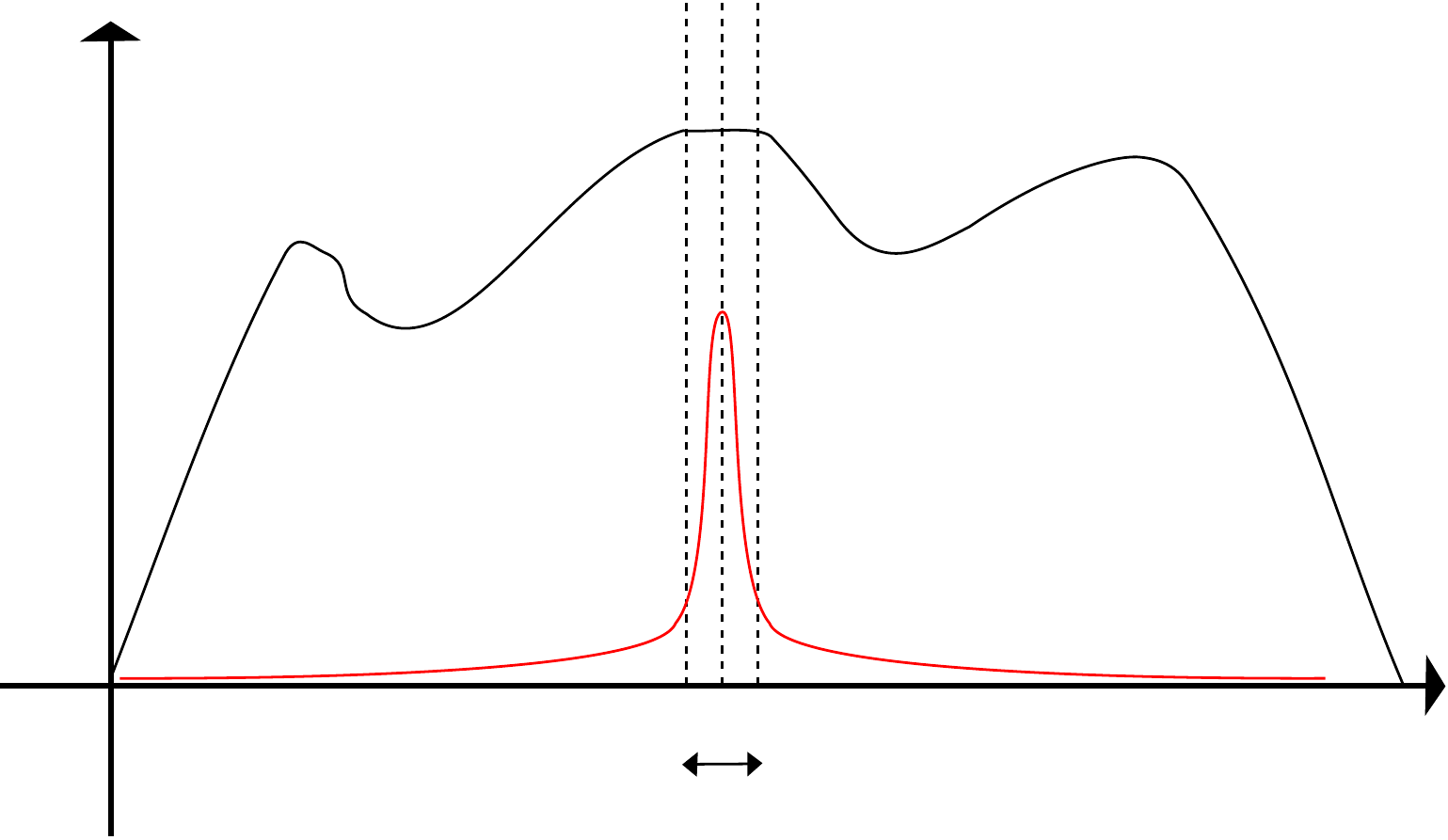}}%
    \put(0.480,0.077){\color[rgb]{0,0,0}\makebox(0,0)[lb]{\smash{$\omega_0$}}}%
    \put(0.95,0.06913388){\color[rgb]{0,0,0}\makebox(0,0)[lb]{\smash{$\omega$}}}%
    \put(0.480,0.00977939){\color[rgb]{0,0,0}\makebox(0,0)[lb]{\smash{$2a$}}}%
    \put(0.167009,0.43){\color[rgb]{0,0,0}\makebox(0,0)[lb]{\smash{$\im \chi(\omega)$}}}%
    \put(0.52,0.27648538){\color[rgb]{0,0,0}\makebox(0,0)[lb]{\smash{{\color{red} $\im L_{+}(\omega, \omega_0)$}}}}%
  \end{picture}%
\endgroup%
\caption{An arbitrary response $\im \chi (\omega)$ to be expressed as a superposition of finite-sized Lorentzians.}
\label{fig:DeltaKonv}
\end{figure}
\begin{eqnarray}
\im &L_{\pm}&( \omega, \omega_0) =\nonumber \\
&&\frac{1}{\pi} \frac{1}{\sqrt{1 - (\Gamma/2\omega_0)^2}} \frac{\frac{1}{2} \Gamma}{\big (\omega \mp \omega_0\sqrt{1 - (\Gamma/2\omega_0)^2} \ \big )^2 + ( \frac{\Gamma}{2} )^2 }.\nonumber \\ \label{eq:LorPeak1}
\end{eqnarray} Introducing \eqref{eq:ExpressLorthruF2} into \eqref{eq:ImChi} (without the limit) then allows the delta-convergence error $\psi_\text{err}(\omega)$ to be expressed as
\begin{eqnarray}
&&\psi_\text{err}(\omega) = \bigg | \int_{-\infty}^\infty \im \chi(\omega_0) \im L_{+}(\omega,\omega_0) \diff \omega_0 - \im \chi(\omega) \bigg | \nonumber \\
&&= \bigg |\int_{-\infty}^\infty  \underbrace{\im \chi \bigg (\omega_0 \sqrt{1 + \frac{\Gamma^2}{4\omega_0^2}} \bigg )}_{=\im \chi_F(\omega_0)} F_\Gamma(\omega_0 -\omega) \diff \omega_0 - \im \chi(\omega) \bigg |. \nonumber \\ \label{eq:CritDeltaConv}
\end{eqnarray} Here, by having made an appropriate substitution, it is observed in the last line that the integral can be written in terms of the Lorentz-Cauchy function $F_\Gamma(\omega_0-\omega)$:
\begin{eqnarray}
F_\Gamma(\omega_0-\omega)= \frac{1}{\pi} \frac{\frac{1}{2} \Gamma}{(\omega_0 -\omega)^2 +  ( \frac{\Gamma}{2} )^2 }. \label{eq:LorentzCauchy}
\end{eqnarray} As a consequence of the substitution, a new function $\im \chi_F(\omega_0)$ is defined in \eqref{eq:CritDeltaConv} which essentially represents a shifted version of $\im \chi(\omega_0)$. 

The integration interval in \eqref{eq:CritDeltaConv} may be divided into three intervals corresponding to the intervals designated in Fig. \ref{fig:DeltaKonv}, which shall then be evaluated separately: 

\begin{eqnarray}
&&\int_{-\infty}^\infty \im \chi_F(\omega_0) F_\Gamma(\omega_0 - \omega) \diff \omega_0 = \nonumber \\
&\bigg (& \underbrace{\int_{-\infty}^{\omega - a}}_{A}  +\underbrace{\int_{\omega - a}^{\omega + a}}_{B} +\underbrace{\int_{\omega + a}^\infty}_{C} \bigg ) \im \chi_F(\omega_0) F_\Gamma(\omega_0 - \omega) \diff \omega_0. \nonumber \\ \label{eq:DivisionOfIntegral}
\end{eqnarray} Here the parameter $a$ is in principle arbitrary, but can be thought to represent the region around $\omega=\omega_0$ where $\im \chi_F(\omega)$ is assumed to be Taylor expandable in the lowest orders. Evaluating integrals (A) and (C) in \eqref{eq:DivisionOfIntegral} together, an upper bound on their sum may be found:
\begin{eqnarray}
&&\max_{\omega'} |\im \chi_F(\omega')| \bigg (\int_{-\infty}^{\omega-a} + \int_{\omega + a}^{\infty} \bigg ) F_\Gamma(\omega_0 - \omega) \diff \omega_0 \nonumber \\
&&= \max_{\omega'} |\im \chi(\omega')| \bigg [ 1 - \frac{2}{\pi} \arctan \bigg ( \frac{2a}{\Gamma} \bigg ) \bigg ] \label{eq:upperBoundSide}
\end{eqnarray} Here it has been used that $\max_{\omega'}|\im \chi_F(\omega')|= \max_{\omega'}|\im \chi(\omega')|$. One observes that as $\Gamma \ll a$ the $\arctan$-term can be expanded, allowing \eqref{eq:upperBoundSide} to be re-expressed as
\begin{eqnarray}
\frac{\Gamma}{\pi a}\max_{\omega'} |\im \chi(\omega')|, \label{eq:expandedSideIntgrls}
\end{eqnarray} which evidently converges to zero when $\Gamma \to 0$ as long as $\im \chi (\omega)$ is bounded. In order to calculate (B), $\im \chi_F(\omega_0)$ is expanded around $\omega_0=\omega$:
\begin{eqnarray}
\im \chi_F (\omega_0) &=& \im \chi_F(\omega) + \im \chi_F ' (\omega) (\omega_0 -\omega) \nonumber \\
&+& \frac{1}{2} \im \chi_F '' (\omega) (\omega_0 - \omega)^2 + O[(\omega_0-\omega)^3] \label{eq:TaylorExpnded}
\end{eqnarray} Due to the parity when inserting this into (B) only even order terms remain, giving
\begin{eqnarray}
\int_{\omega - a}^{\omega + a} \im \chi_F(&\omega_0&) F_\Gamma(\omega_0 - \omega) \diff \omega_0   \nonumber \\
&=&  \frac{2}{\pi} \im \chi_F (\omega) \arctan \bigg ( \frac{2 a}{\Gamma} \bigg )  \nonumber \\
&&  + \frac{\Gamma a}{2 \pi} \im \chi_F '' (\omega) + O \bigg ( \frac{\Gamma}{ a}  \bigg ).  \label{eq:seriesExpResult}
\end{eqnarray} Only $\Gamma$-terms to the first order have been kept in arriving at this expression. The term $O(\Gamma/a)$ corresponds to terms containing higher order derivatives of $\im \chi_F(\omega)$, which will be assumed to be negligible. In the event that there exists significant higher order derivatives (e.g. as will be the case in Fig. \ref{fig:Example1} near the steep drop), one may keep more terms in going from \eqref{eq:TaylorExpnded} to \eqref{eq:seriesExpResult} until the next even ordered derivative is negligible. The same steps that now follow may then be applied in order to find the relevant error estimate.

\begin{figure}[htb]
\centering
\includegraphics[width=0.44\textwidth]{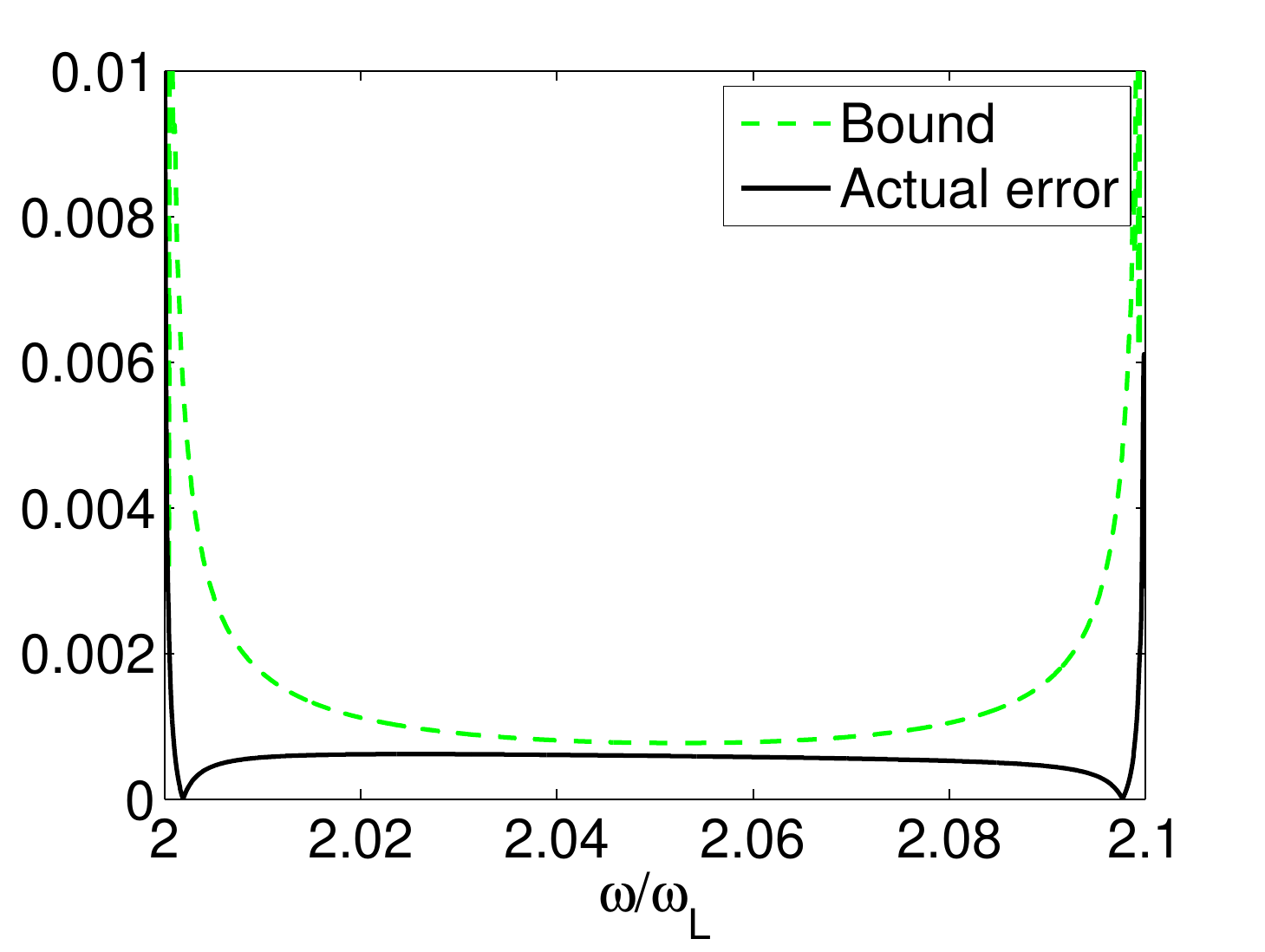}
\caption{The error in approximating the perfect lens response described in Sec. \ref{sec:perfectlense} with a Lorentzian superposition is shown to be bounded by \eqref{eq:Error}.}
\label{fig:PerfLensErrorBound}
\end{figure}

In order to arrive at an expression for $\psi_\text{err}(\omega)$ in terms of the goal function $\im \chi(\omega)$ and its derivatives, the shifted function $\im \chi_F(\omega)=\im \chi(\omega\sqrt{1 + \Gamma^2/4\omega^2})$ is expanded under the assumption that $\Gamma / \omega \ll 1$ and then inserted for $\im \chi_F(\omega)$ and its second derivative in \eqref{eq:seriesExpResult}. Expanding $\arctan$ under the assumption that $\Gamma \ll a$ permits further simplification. Combining the resulting expression with \eqref{eq:expandedSideIntgrls}, yields an upper bound on the delta-convergence error
\begin{eqnarray}
\psi_\text{err}(\omega) \leq \frac{\Gamma}{\pi} \bigg ( \max_{\omega'} |\im \chi(\omega')| \frac{2}{a} &+ &|\im \chi''(\omega)| \frac{a}{2} \bigg ) \nonumber \\ 
&+& O  (\Gamma^2 ),   \label{eq:deltaConvErr}
\end{eqnarray} where $O(\Gamma^2)$ arises after having expanded and replaced $\im \chi_F(\omega)$ and its second derivative. Minimizing the upper bound with respect to $a$ then gives
\begin{equation}
a^2= \frac{4 \max_{\omega'} |\im \chi(\omega')|}{|\im \chi''(\omega)|} . \label{eq:expressionfora}
\end{equation}  An inverse relationship between $a^2$ and $|\im \chi''(\omega)|$ is intuitive given that $a$ must be small when $\im \chi(\omega)$ varies steeply.  Inserting \eqref{eq:expressionfora} into \eqref{eq:deltaConvErr} while neglecting $O(\Gamma^2)$ gives
\begin{eqnarray}
\psi_\text{err}(\omega) \leq \frac{2 \Gamma}{\pi} \bigg (  |\im \chi''(\omega)| \cdot \max_{\omega'} |\im \chi(\omega')| \bigg )^\frac{1}{2}. \label{eq:Error}
\end{eqnarray} Hence, the delta-convergence error $\psi_\text{err}(\omega)$ is proportional to $\Gamma \sqrt{|\im \chi ''(\omega)|}$ under the assumption that $\Gamma \ll a \propto 1/\sqrt{|\im \chi '' (\omega)|}$, the higher order derivatives are negligible, and $\Gamma / \omega \ll 1$. Fig. \ref{fig:PerfLensErrorBound} shows \eqref{eq:Error} applied to the imaginary part of the perfect lens response displayed in Fig. \ref{fig:PerflensSum}, where one observes that the actual error (black curve) is bounded by \eqref{eq:Error} (dashed curve). The displayed error is actually taken to be the total error $|\im \chi(\omega) - \im \chi_{\Delta, \Gamma}(\omega)|$ of the superposition, rather than the difference between $\im \chi(\omega)$ and $\im \chi_\Gamma(\omega)$. However, here $\Delta/\omega_L$ in the sum \eqref{representionsum2} has been reduced by a factor 2 as compared to in Fig. \ref{fig:PerflensSum}, in order to reduce the significance of the Riemann sum error \eqref{errestimation} in comparison to the delta-convergence approximation error \eqref{eq:CritDeltaConv}.


\section{Conclusion}
It has been demonstrated that superpositions of Lorentzian functions are capable of approximating all complex-valued functions satisfying the Kramers-Kronig relations, to any desired precision. The typical text-book analysis of dielectric or magnetic dispersion response functions in terms of Lorentzians is thereby extended to cover the whole class of causal functions. The discussion started with approximating the imaginary part of an arbitrary susceptiblity $\chi(\omega)$ by a superposition of imaginary parts of Lorentzian functions. The error $|\im \chi(\omega) - \im \chi_{\Gamma, \Delta}(\omega)|$ was then shown to become arbitrarily small as $\Gamma \to 0$, $\omega_M \to \infty$ and $\Delta= \Gamma^2 / \omega_M$. Since this error was shown to vanish in $L^2$ it was known that $\re \chi_{\Gamma, \Delta}(\omega) \to \re \chi(\omega)$. From this the superposition $\chi_{\Gamma, \Delta} (\omega)$ was found, which approximates $\chi(\omega)$ to any desired precision.

The superposition $\chi_{\Gamma, \Delta} (\omega)$ has been demonstrated to reconstruct two response functions that obey the Kramers-Kronig relations while not resembling typical Lorentzian resonance behavior. The first response is known to lead to significant negative real values of the susceptibility $\chi(\omega)$ in a narrow bandwidth with arbitrarily low gain or loss, while the other response represents the optimum realization of a perfect lens on a bandwidth. These examples demonstrate the possibility of viewing Lorentzians as useful building-blocks for manufacturing desired responses that are not found in conventional materials. To this end, the ability to realize Lorentzian sums using split ring cylinders with varying parameters was shown. In order to quantify the precision of a superposition approximation, error estimates have been derived.

\appendix
\section{Delta-Convergence of Lorentzians}
\label{sec:DeltaConv}
This section will prove \eqref{eq:LorentzToTwoDeltas} and equivalently \eqref{eq:ImChi}. In doing so, observe that an implicit definition of the delta-function is
\begin{eqnarray}
\im \chi(\omega) &=& \int_{-\infty}^\infty \im \chi(\omega_0) \delta(\omega_0 -\omega) \diff \omega_0 \nonumber \\
&=& \int_0^\infty \im \chi(\omega_0) \big [ \delta(\omega_0 -\omega) - \delta(\omega_0 + \omega) \big ] \diff \omega_0.  \nonumber \\
\end{eqnarray} Note that the last line holds only insofar as the goal function obeys odd symmetry $\im \chi(-\omega_0)= -\im \chi (\omega_0)$.  By inserting \eqref{eq:ExpressLorthruF2} into \eqref{eq:ImChi} and using the symmetry $L_{-}(\omega, -\omega_0)= L_{+}(\omega, \omega_0)$, one observes that the goal of this section becomes to demonstrate that



\begin{eqnarray}
\underset{\Gamma \to 0}{\lim} \int_{-\infty}^\infty \im \chi(\omega_0) \im L_{+}(\omega,\omega_0) \diff \omega_0 = \im \chi(\omega).
\label{eq:DefDeltaConv}
\end{eqnarray} Inserting for $\im L_{+}(\omega,\omega_0)$ in the integral and making the appropriate substitution leads to:
\begin{equation}
\lim_{\Gamma \to 0} \frac{1}{\pi} \int_{-\infty}^{\infty} \frac{\im \chi(\frac{\Gamma}{2}\sqrt{(2 \omega /\Gamma - u)^2 + 1})}{u^2 + 1} \diff u. \label{eq:substImChi}
\end{equation} One may then move the limit inside the integral under the criteria of Lebesgue's Dominated Convergence Theorem. As $\Gamma \to 0$ the function $\im \chi(\frac{\Gamma}{2}\sqrt{(2 \omega /\Gamma - u)^2 + 1})$ converges pointwise to $\im \chi(\omega)$. Furthermore, it is clear that an integrable function that bounds the integrand for all values of $\Gamma$ exists under the condition that $\im \chi(\omega)$ is bounded. Hence, \eqref{eq:substImChi} gives
\begin{equation}
\frac{1}{\pi} \im \chi(\omega) \int_{-\infty}^{\infty} \frac{1}{u^2 + 1} \diff u \label{eq:LimsubstImChi}= \im \chi(\omega),
\end{equation} which concludes the proof.

\section{L2 convergence} 
\label{sec:L2Conv}
This section considers the conditions upon $\Gamma$, $\Delta$ and $\omega_M$ needed in order to obtain the $L^2$ convergence expressed in \eqref{eq:L2ImChiNew}. From the discussion in Sec. \ref{sec:Errorestimate}, one may express:


\begin{eqnarray}
&\bigg |&\im \chi(\omega)- \im \chi_{\Gamma, \Delta}(\omega)  \bigg | \leq \nonumber \\
&& \bigg | \im \chi(\omega)  \frac{\omega_M \Delta^2}{3 \Gamma^3} \bigg ( 1+ O \bigg ( \frac{\Gamma^2}{\omega^2} \bigg ) \bigg ) \bigg | \label{eq:L2conv1}\\
&+& \bigg | \frac{2}{\pi} \int_{\omega_M}^\infty \im \chi (\omega_0) \im \bigg \{ \frac{\omega_0}{\omega_0^2 - \omega^2 - i \omega \Gamma} \bigg \} \diff \omega_0 \bigg | \label{eq:L2conv2} \\
&+& \bigg | \bigg ( \frac{2}{\pi} \arctan \bigg (\frac{2 a}{\Gamma} \bigg ) - 1 \bigg ) \im \chi(\omega) \bigg | \label{eq:L2conv3} \\
&+& \bigg | \frac{\Gamma a}{2 \pi} \im \chi '' (\omega) + O(\Gamma^2) \bigg | \label{eq:L2conv4} \\
&+& \bigg | \bigg ( \int_{-\infty}^{\omega - a} + \int_{\omega + a}^{\infty} \bigg ) \im \chi_F (\omega_0) F_\Gamma(\omega_0-\omega) \diff \omega_0 \bigg | \label{eq:L2conv5}
\end{eqnarray} The Riemann sum error \eqref{errestimation} is expressed through \eqref{eq:L2conv1} and \eqref{eq:L2conv2}, and the delta-convergence error from \eqref{eq:DivisionOfIntegral} and \eqref{eq:seriesExpResult} is expressed in \eqref{eq:L2conv3}, \eqref{eq:L2conv4} and \eqref{eq:L2conv5}. The term $O(\Gamma^2)$ in \eqref{eq:L2conv4} is included due to having expanded and replaced $\im \chi_F(\omega)$ and its second derivative so as to express \eqref{eq:seriesExpResult} in terms of $\im \chi(\omega)$ and its second derivative. 

The bound given by \eqref{eq:L2conv1}-\eqref{eq:L2conv5} has been derived under the assumption that $\omega \gg \Gamma$. Before proceeding to evaluate the $L^2$ norm by these, one must therefore demonstrate that $|\im \chi(\omega)- \im \chi_{\Gamma, \Delta}(\omega)|$ is bounded in a region $|\omega| \in (0, \beta)$ where $\beta \gg \Gamma$ (representing the region not accounted for by \eqref{eq:L2conv1}-\eqref{eq:L2conv5}), so that as $\Gamma \to 0$ and then $\beta \to 0$, it is known that the $L^2$ norm converges to zero also here. A detailed proof of this is will not be given here, but its result can be understood by noting that \eqref{eq:ImChi}, when without its limit, is bounded by an integral of $\im L(\omega)$ on the interval $\omega_0 \in (0, \infty)$ multiplied with $\max_{\omega'}|\im \chi(\omega')|$. Since the integral is finite and $\chi(\omega)$ is analytic, it follows that \eqref{eq:ImChi} is bounded. Furthermore, since the sum \eqref{representionsum} can be made arbitrarily close to \eqref{eq:ImChi}, it is intuitive that it also remains bounded for all $\omega$. Since $\im \chi(\omega)$ and $\im \chi_{\Gamma, \Delta}(\omega)$ are bounded it follows that $|\im \chi(\omega)- \im \chi_{\Gamma, \Delta}(\omega)|$ is finite for all $\omega$. 

Proceeding now to evaluate the $L^2$ norm from \eqref{eq:L2conv1}-\eqref{eq:L2conv5}, convergence for \eqref{eq:L2conv1} is achieved by setting $\Delta= \Gamma^2/\omega_M$ and taking the limit $\Gamma \to 0$, since $\im \chi(\omega)$ is in $L^2$ (the presence of the term $O(\Gamma^2/\omega^2)$ will be discussed below). The same convergence occurs in \eqref{eq:L2conv3} as $\Gamma \to 0$. The $L^2$ norm arising for \eqref{eq:L2conv4}


\begin{equation}
\frac{\Gamma a}{2 \pi}\big \| \im \chi ''(\omega) \big \|  , \label{eq:L2conv4.2} 
\end{equation} converges provided that $\im \chi''(\omega)$ is in $L^2$. Since it however is conceivable for a function $\chi(\omega)$ in $L^2$ to have derivatives not in $L^2$, one may instead consider $\chi(\omega)$ along the line $\omega + i \delta$, where $\delta >0$ is an arbitrarily small parameter. One then finds through the Fourier transform
\begin{eqnarray}
\chi( \omega + i \delta) &=& \int_0^\infty x(t) \text{e}^{i(\omega + i \delta)} \diff t \nonumber \\
&=& \int_0^\infty [x(t)\text{e}^{-\delta t}]  \text{e}^{i\omega t} \diff t, \label{eq:FourierIntegral}
\end{eqnarray} given that $x(t)$ is the time-domain response associated with $\chi(\omega)$. Eq. \eqref{eq:FourierIntegral} reveals that one may express
\begin{eqnarray}
\chi(\omega + i \delta) &\xrightarrow{\mathscr{F}^{-1}}& x(t) \text{e}^{-\delta t} \nonumber \\
\chi''(\omega + i \delta) &\xrightarrow{\mathscr{F}^{-1}}& (it)^2 x(t) \text{e}^{-\delta t}. \label{eq:TransformIdentities}
\end{eqnarray} Since $x(t)$ is in $L^2$ one observes that $(it)^2 x(t) \text{e}^{-\delta t}$ is in $L^2$ for $\delta >0$. Knowing that the Fourier transform preserves the $L^2$ norm, it follows that $\chi''(\omega + i \delta)$ is in $L^2$.

The terms $O(\Gamma^2/\omega^2)$ and $O(\Gamma^2)$ in \eqref{eq:L2conv1} and \eqref{eq:L2conv4} involve multiples of either $\im \chi(\omega)$ or its derivatives. From \eqref{eq:TransformIdentities} it is clear that derivatives of all orders are assured to be in $L^2$ by the above procedure. Hence these terms vanish as $\Gamma \to 0$. Note that in the derivation of $\eqref{eq:L2conv4}$, derivatives greater than the second order of $\im \chi(\omega)$ have been neglected, as discussed with regards to \eqref{eq:seriesExpResult}. If these were to be included here, they would lead to terms of the same form as \eqref{eq:L2conv4.2} and would converge in the same manner.



Considering now \eqref{eq:L2conv2}, one observes that by using \eqref{eq:ExpressLorthruF2} one may express this as:
\begin{eqnarray}
\bigg | \bigg ( \int_{-\infty}^{-\omega_M} + \int_{\omega_M}^{\infty} \bigg ) \im \chi (\omega_0) \im L_{+}(\omega,\omega_0) \diff \omega_0 \bigg |.
\end{eqnarray} To find the limit of the $L^2$ norm, the task here is therefore to evaluate
\begin{eqnarray}
\lim_{\Gamma \to 0} \int_{-\infty}^\infty\bigg | \bigg (\int_{-\infty}^{-\omega_M} + \int_{\omega_M}^\infty \bigg ) \im \chi (\omega_0) \im L_{+}(\omega,\omega_0) \diff \omega_0 \bigg |^2 \diff \omega . \nonumber \\ \label{eq:L2normIntForm}
\end{eqnarray}Moving the limit inside the outermost integral is permitted through Lebesgue's Dominated Convergence Theorem, under the condition that $\im \chi (\omega_0)$ is bounded. In the event that $\im \chi (\omega_0)$ is singular on the real axis, one may instead use $\im \chi (\omega_0 + i \delta)$ (where $\delta>0$ is an arbitrarily small parameter) for which any divergence is quenched, as discussed in the introduction. This gives
\begin{eqnarray}
&&\int_{-\infty}^\infty \bigg | \lim_{\Gamma \to 0} \bigg (\int_{-\infty}^{-\omega_M} + \int_{\omega_M}^\infty \bigg ) \nonumber \\
&& \quad \quad \quad \quad \quad \quad \im \chi (\omega_0) \im L_{+}(\omega,\omega_0) \diff \omega_0 \bigg |^2 \diff \omega \nonumber \\ 
&&= \bigg (\int_{-\infty}^{-\omega_M} + \int_{\omega_M}^\infty \bigg ) |\im \chi (\omega)|^2 \diff \omega, \label{eq:L2OmM3}
\end{eqnarray} when having used \eqref{eq:DefDeltaConv}. The final result \eqref{eq:L2OmM3} reveals that the norm is finite, and that one however must demand $\omega_M \to \infty$ in order that \eqref{eq:L2conv2} converges to zero.


The remaining task is to evaluate \eqref{eq:L2conv5} through the limit
\begin{eqnarray}
\lim_{\Gamma \to 0} \int_{-\infty}^\infty\bigg |\bigg ( \int_{-\infty}^{\omega-a} + \int_{\omega + a}^{\infty} \bigg ) \im \chi_F (\omega_0) F_\Gamma(\omega_0-\omega) \diff \omega_0 \bigg |^2 \diff \omega. \nonumber \\ \label{eq:OmegaAIntegralL2}
\end{eqnarray} Under the same condition as before one may move the limit inside the outermost integral through Lebesgue's Dominated Convergence Theorem:
\begin{eqnarray}
\int_{-\infty}^\infty &\bigg |& \lim_{\Gamma \to 0} \bigg ( \int_{-\infty}^{\omega-a} + \int_{\omega + a}^{\infty} \bigg ) \im \chi_F (\omega_0) F_\Gamma(\omega_0 - \omega) \diff \omega_0 \bigg |^2 \diff \omega =0. \nonumber \\ 
\end{eqnarray} Here the delta-convergence property of the Lorentz-Cauchy function $F_\Gamma(\omega_0-\omega)$  has been used.

Hence, it has been demonstrated that  the $L^2$ convergence expressed in \eqref{eq:L2ImChiNew} is achieved by setting $\Gamma \to 0$, $\omega_M \to \infty$ and $\Delta= \Gamma^2/\omega_M$.

\def\cprime{$'$}

\end{document}